\documentclass[preprint,showpacs]{revtex4}
\usepackage{graphicx}
\begin{document}
\title{Noise in Disordered Systems: Higher Order Spectra in Avalanche Models } 
\author{ Amit P. Mehta }
\email[]{apmehta@uiuc.edu}
\affiliation{ Department of Physics, University of Illinois at
Urbana-Champaign, 1110 West Green Street, Urbana, IL 61801-3080 }
\author{Karin A. Dahmen}
\email[]{dahmen@uiuc.edu}
\affiliation{ Department of Physics, University of Illinois at
Urbana-Champaign, 1110 West Green Street, Urbana, IL 61801-3080 }
\author{M. B. Weissman}
\email[]{mbw@uiuc.edu}
\affiliation{ Department of Physics, University of Illinois at
Urbana-Champaign, 1110 West Green Street, Urbana, IL 61801-3080 }
\author{Tim Wotherspoon}
\affiliation{ Department of Physics, University of Illinois at
Urbana-Champaign, 1110 West Green Street, Urbana, IL 61801-3080 }

\date{\today} 
\newcommand{\summ}{\sum_{m=1}^{j-1}}
\newcommand{\be}{\begin{equation}}
\newcommand{\ee}{\end{equation}}
\newcommand{\bea}{\begin{eqnarray*}}
\newcommand{\eea}{\end{eqnarray*}}

\begin{abstract}

        We present a novel analytic calculation of the Haar power
        spectra, and various higher order spectra, of mean field
        avalanche models. We also compute these spectra from a
        simulation of the zero-temperature mean field RFIM and
        infinite range RFIM model for $d = 3$. We compare the results
        and obtain novel exponents.
        
\end{abstract}


\maketitle
        {\em Introduction}: Certain disordered systems {\em crackle},
        that is when such a system is slowly driven it responds by
        discrete jumps or avalanches of a broad range of sizes \cite{Nature}.
        This crackling is often the result of non-equilibrium
        collective transport in random media, where the interactions
        are strong enough that thermal effects are negligible and the
        system has many interacting degrees of freedom. Such systems
        include: charge density waves, vortices in type II
        superconductors, crack propagation, earthquakes, and
        Barkhausen noise in magnets.  In mean field theory the
        avalanche dynamics of the systems mentioned above are
        described by Possion statistics (given by Eq. (\ref{prodist}))
        at the critical point or critical depinning transition
        \cite{Matt1,Fisher}.
        
        In our novel mean field analysis we determine the following
        spectral functions: Haar power spectra, $1.5$ spectra, and
        second spectra for systems that have avalanche dynamics given
        by Eq. (\ref{prodist}). Haar power spectra allow us to obtain
        a power series in time, $H(t,f_1)$, needed to calculate higher
        order spectra.  Higher order spectra give valuable information
        about the avalanche dynamics in Barkhausen noise not
        accessible through ordinary power spectra \cite{OBrien,Petta}.
        Also, higher order spectra have been used to obtain crucial
        information about a variety of diverse systems such as:
        metastable states in vortex flow \cite{Merithew}, natural
        auditory signals \cite{Thomson}, conductance-noise power
        fluctuations \cite{Parsam}, non-Gaussian $1/f$ noise
        \cite{Seidler}, and spin glasses \cite{Weissman}. While much
        experimental work has been done studying higher order spectra
        \cite{OBrien,Petta,Merithew,Thomson,Parsam,Seidler,Weissman},
        we present a rigorous mean field treatment that is applicable
        to a broad range of systems \cite{Fisher}. This analysis is
        one of a small number of theoretical calculations in higher
        order noise statistics.


        We also compare our general results from mean field theory to
        a mean field simulation of the $T=0$ random field Ising model
        (RFIM) \cite{Dahmen}, and an analysis of Barkhausen noise
        obtained from a simulation of the $T=0$ infinite range RFIM in
        $d=3$. We also find novel exponents from our analysis.

        {\em Mean Field RFIM}: The $T=0$ mean field RFIM consists of
        an array of $N$ spins ($s_i = \pm 1$), which may point up
        ($s_i = +1$) or down ($s_i = -1$). Spins are coupled to all
        other spins (through a ferromagnetic exchange interaction
        $J$), and to an external field $H(t)$ which is increased
        adiabatically slowly.  To model dirt in the material, we
        assign a random field, $h_i$, to each spin, chosen from a
        distribution $P(h_i) = exp(-h^2_i/2R^2)/\sqrt{2\pi}R$, where
        $R$, the disorder, determines the width of the Gaussian
        probability distribution and therefore gives a measure of the
        amount of quenched disorder for the system.The Hamiltonian for
        the system at a time $t$ is given by: $H = -\sum\limits_i(JM +
        H(t) + h_i)s_i$, where $M = \frac{1}{N}\sum\limits_j s_j$ is
        the magnetization of the system. Initially, $H(-\infty) =
        -\infty$ and all the spins are pointing down. Each spin is
        always aligned with its local effective field $h_i^{eff} = JM
        + H(t) + h_{i}$.

        {\em RFIM with Infinite Range Forces in 3D}: The $T=0$ Infinite
        Range RFIM (IRM) consists of a 3D lattice of $N$ spins, with a
        Hamiltonian at a time $t$ is given by: $H = \sum\limits_{<ij>}
        -J s_i s_j - \sum_{i}(H(t) + h_i - J_{inf}M)s_i$, where
        $J_{inf}>0$ is the strength of the infinite range demagnetizing
        field, and $\langle ij\rangle$ stands for nearest neighbor
        pairs of spins. The local effective field is given by
        $h_i^{eff} = J\sum_{<ij>}s_j + H(t) + h_{i} - J_{inf}M$.
        
        The addition of a weak $J_{inf} \sim \frac{1}{N}$ to the
        traditional RFIM causes the system to exhibit self-organized
        criticality (SOC) \cite{Urbach,Narayan,Zapperi}.  This means
        that as $H$ is increased the model always operates at the
        critical depinning point, and no parameters need to be tuned
        to exhibit critical scaling behavior (except
        $\frac{dH}{dt}\rightarrow 0$).  We limit our analysis to a
        window of $H$ values where the slope of $M(H)$ is constant and
        the system displays front propagation behavior.  Details of
        the simulation are given elsewhere \cite{Matt2}.

        {\em Avalanche Dynamics in the $T=0$ RFIM}: The external field
        $H(t)$ is adiabatically slowly increased from $-\infty$ until
        the local field, $h_i^{eff}$, of any spin $s_i$ changes sign,
        causing the spin to flip \cite{Matt1,Sethna}.  It takes some
        microscopic time $\Delta t$ for a spin to flip. 
        The spin flip changes the
        local field of the coupled spins and may cause them to flip as
        well, etc.  This {\em avalanche} process continues until no
        more spin flips are triggered.  Each step of the avalanche,
        that is each $\Delta t$, in which a set of spins
        simultaneously flip, is called a {\em shell}. The number of
        spins that flip in a shell is directly proportional to the
        voltage $V(t)$ during the interval $\Delta t$ that an
        experimentalist would measure in a pick-up coil wound around
        the sample. In our simulations we denote the number of spins
        flipped in a shell at a time $t$ by $n_t (= V(t))$.  The first
        shell of an avalanche (one spin flip) is triggered by the
        external field $H(t)$, while each subsequent shell within the
        avalanche is triggered only by the previous shell, since
        $H(t)$ is kept constant while the avalanche is propagating.
        $H(t)$ is only increased when the current avalanche has
        Stopped, and is increased only until the next avalanche is
        triggered (i.e. $\frac{dH}{dt}\rightarrow 0$).  The number of
        shells in an avalanche times $\Delta t$ defines the {\em pulse
        duration}, $T$, or the time it took for the entire avalanche
        to flip.  The time series of $n_t$ values for many successive
        avalanches creates a Barkhausen train.

        {\em Poisson Distribution}: In order to calculate the Haar
        power spectrum and higher order spectra we first need the
        probability distribution for an (infinite) avalanche at
        criticality in the class of mean field avalanche models we are
        studying \cite{Fisher}, which is given by the following
        Poisson distribution \cite{Fisher,Matt1}:

\begin{equation} \label{prodist}
        P(n_0 = 1,n_1,n_2,...,n_{\infty}) = \frac{1}{e n_1!}
        \prod\limits^{\infty}_{t=2}\frac{e^{-n_{t-1}}n^{n_t}_{t-1}}{n_t!}
\end{equation}  

An avalanche begins with a discrete jump of unit magnitude, so we have
$n_0 = 1$, in the context of Barkhausen noise each avalanche begins
with one spin flip. Now
$\langle...\rangle\equiv\sum\limits_{\{n_1,...,n_{\infty}\}}...
P(n_1,...n_{\infty})$ represents the average over Eq. (\ref{prodist}).
Using Eq. (\ref{prodist}) we determine the 2-pt, 3-pt, and 4-pt
time-time correlation functions, that allow us to analytically
calculate the Haar spectra and higher order spectra. Details are left
for a long paper.

        
{\em Haar Series and Haar Transform}: The Haar transform, $H(t,f_1)$,
a simple wavelet transformation, is the square wave equivalent of the
Fourier transform; where $H(t,f_1)$ is the absolute square of the time
integral over a period (of duration $\frac{1}{f_1}$) of the square
wave times a section of the train centered around $t$ \cite{OBrien}.
The Haar transform affords us the needed time resolution at the
expense of frequency resolution that is extraneous to our purpose,
since we are interested in studying how the power contribution at a
given frequency $f_1$ changes in time.
        
        We first determine the Haar power series for a single large
        avalanche of duration $T$ (in the limit that $T \rightarrow
        \infty$) using the 2-pt correlation function.  We obtain the
        following exact result: $H(t,f_1) =
        \frac{1}{12}(2\frac{(\Delta t)^2}{f_1} + \frac{1}{f_1^3})$
        Moreover, to find the Haar power we sum our result over the
        whole avalanche to obtain:

\begin{equation} \label{calc2}
        S_H(f_1) \equiv \frac{1}{T}\sum_{t=1}^{Tf_1} H(t,f_1) =
        \frac{1}{12}(2(\Delta t)^2 + \frac{1}{f_1^2}) 
\end{equation}

This result differs from the Fourier power, $S_F(f_1) =
\frac{1}{f_1^2}$ \cite{Matt1}, by an additive constant and a constant
factor.  This is an artifact of doing a discrete transform. We sum
from $t=1$ to $Tf_1$ (not to $T$) since we are summing over Haar
wavelets of duration $\frac{1}{f_1}$. We reaffirm the above result
(Eq. (\ref{calc2})) through a mean field simulation.


        {\em $1.5$ Spectra and Second Spectra}: The $1.5$ spectra,
        and second spectra are defined below:

\begin{eqnarray} \label{halfspec}
&& S_{1.5}(f_2,f_1) = \frac{\langle
F_t\{v(t,f_1)\}F_t^*\{H(t,f_1)\}\rangle}{\sum_t H(t,f_1)} \\
\label{secondspec}
&& S_2(f_2,f_1) = \frac{\langle
F_t\{H(t,f_1)\}F_t^*\{H(t,f_1)\}\rangle}{(\sum_t H(t,f_1))^2} 
\end{eqnarray}

        where $v(t,f_1)$ is the sum of $n_i$ around time $t$ over a
        duration $\frac{1}{f_1}$, and $f_2$ is the frequency conjugate
        to $t$. Also, $F_t\{...\}$ is the discrete Fourier transform (FT)
        with respect to $t$, and $\sum_t \equiv\sum_{t=1}^{Tf_1}$.
        
\begin{figure}
\includegraphics[scale= .45]{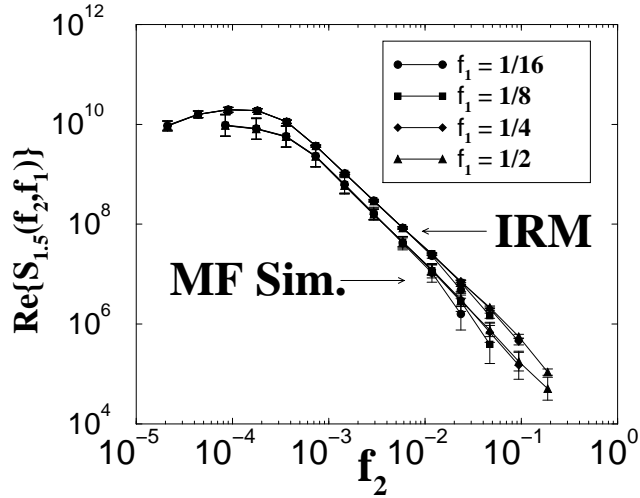}
\caption{\label{halfspecplot} We present the $Re\{S_{1.5}(f_2,f_1)\}$ in the MF
  Sim. and in the IRM. At high frequency $Re\{S_{1.5}(f_2,f_1)\}\sim
  f_2^{-V_{1.5}}$ (go to Table \ref{expo}). There is no visible
  flattening present due to $\Gamma_{1.5}^{(N)}(f_1)$ ($\theta=0$)
  term, since the magnitude of this term is small relative to the
  $f_2$ dependent term. }
\end{figure}

\begin{figure}
\includegraphics[scale= .45]{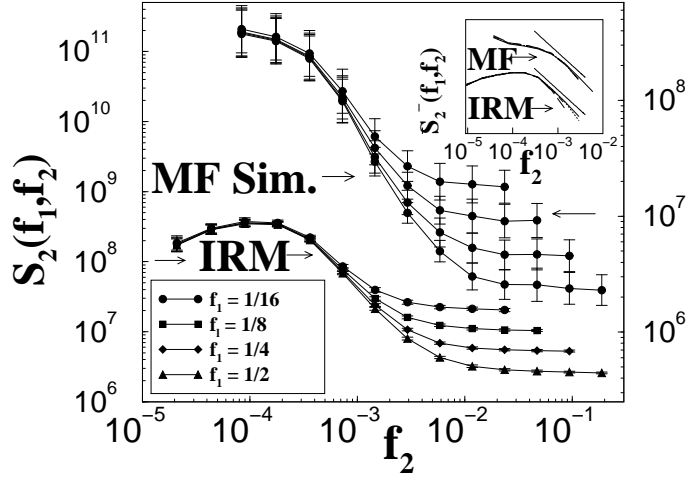}
\caption{\label{secondspecplot} We present the $S_2(f_2,f_1)$ in the
  MF Sim. and IRM. Notice the flattening due to the
  $\Gamma_2^{(N)}(f_1)$ ($\theta=0$) term. At high frequency
  $S_2(f_2,f_1)\sim f_2^{-V_2}+$ (DC background). We use a
  non-linear curve fit of the form: $A0*x^{-A1}+A2$ to determine the
  $V_2$ exponent given in Table \ref{expo}. Inset: $S_2^-(f_2,f_1)$ is
  the second spectra in the MF Sim. and the IRM with the theoretical
  DC background term subtracted. The bold lines adjacent to the
  MF curve and the IRM curve have an exponent of -2 and -1.8,
  respectively.  Please note that these curves (in the inset) do not
  show the full range of data points, and are simply presented to
  illustrate that we have a power law in the absence of the DC
  background term. }
\end{figure}

        To calculate Eq. (\ref{halfspec}-\ref{secondspec}) we first
        write the product of FTs as the FT of a convolution. This
        leaves us with the following quantities: $\langle
        v(t,f_1)H(t+\theta,f_1)\rangle$, and $\langle
        H(t,f_1)H(t+\theta,f_1)\rangle$ where $\theta$ is the
        convolution variable.  These quantities may then be rewritten
        as sums of 3-pt or 4-pt correlation functions, and
        subsequently evaluated.  We obtain the following results:

\begin{eqnarray} \label{theta1}
&&\Gamma_{1.5}(f_1)\equiv \sum_{t=1}^{Tf_1}\langle v(t,f_1)H(t,f_1)\rangle \simeq
\frac{A_{1.5}}{f_1^{Q_{1.5}}}\\
\label{theta2}
&&\Gamma_2(f_1)\equiv \sum_{t=1}^{Tf_1}\langle H(t,f_1)H(t,f_1)\rangle \simeq
        \frac{A_2}{f_1^{Q_2}}\\
\label{halfSC}
&& Re\{S_{1.5}(f_2,f_1)\} = \frac{B_{1.5}}{f_2^{V_{1.5}}}+ \Gamma_{1.5}^{(N)}(f_1)\\
\label{secondSC}
&& S_2(f_2,f_1) =  \frac{B_2}{f_2^{V_2}} + \Gamma_2^{(N)}(f_1)
\end{eqnarray}

where $V_{1.5}$, $V_2$, $Q_{1.5}$, $Q_2$ are given in Table \ref{expo}
for mean field theory, the mean field simulation, and the IRM. Also,
$A_{1.5}$, $A_2$, $B_{1.5}$, and $B_2$ are non-universal constants.
Plots of Eqs. (\ref{halfSC}-\ref{secondSC}) are given in Figs.
(\ref{halfspecplot})-(\ref{secondspecplot}).

\begin{figure}
\includegraphics[scale= .45]{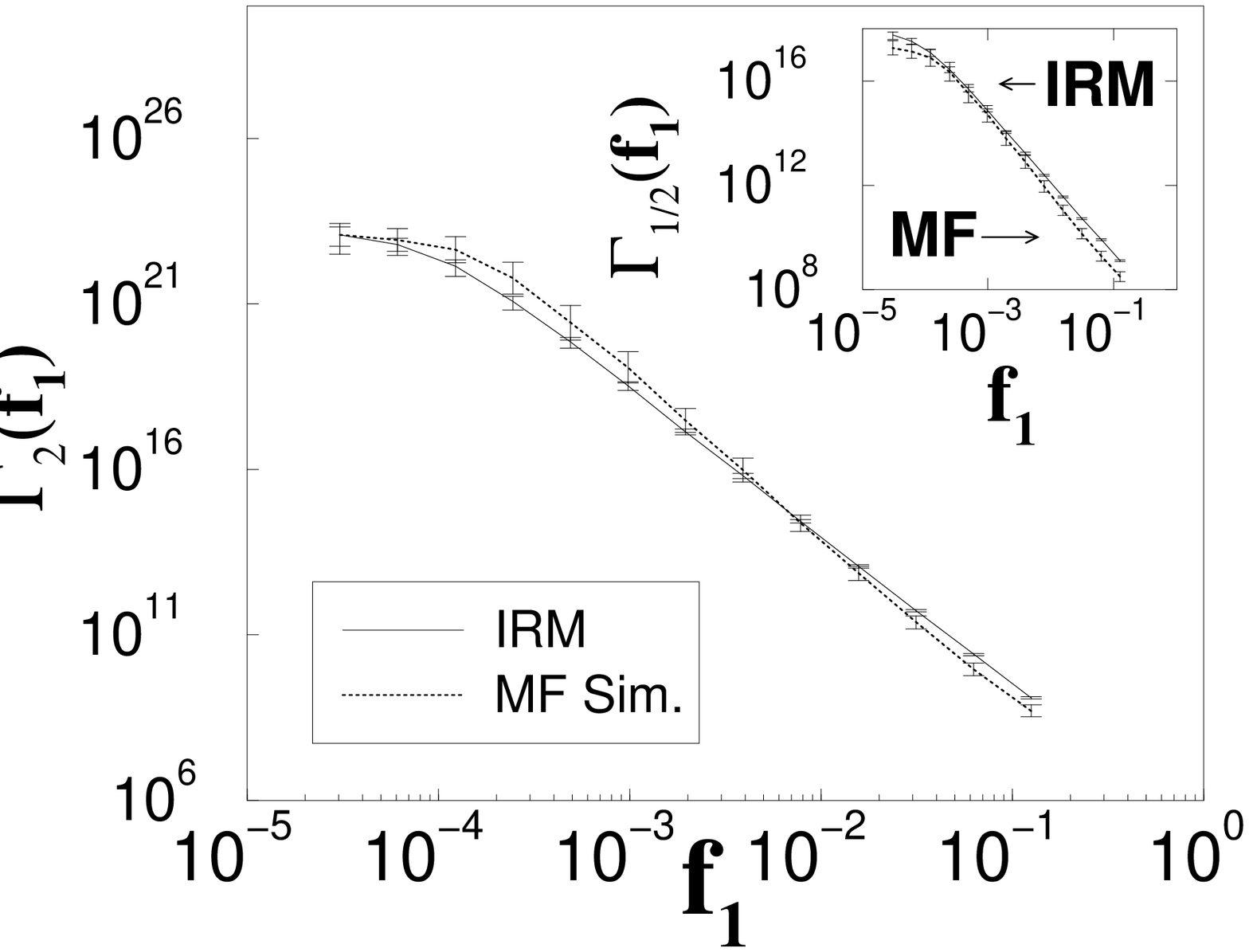}
\caption{\label{haarsumplot} We present $\Gamma_2(N)$ in the MF
Sim. and in the IRM. For high frequency $\Gamma_2(N)\sim
f_1^{-Q_2}$. Inset: $\Gamma_{1/2}(N)$ for the MF
and the IRM. For high frequency $\Gamma_{1/2}(N)\sim f_1^{-Q_{1.5}}$,
see Table \ref{expo}.}
\end{figure}

Now $\Gamma_{1/2}(f_1)$ and $\Gamma_2(f_1)$ are independent of $f_2$,
and result from the $\theta=0$ evaluation of $\langle
v(t,f_1)H(t+\theta,f_1)\rangle$, and $\langle
H(t,f_1)H(t+\theta,f_1)\rangle$ (see Fig. \ref{haarsumplot}). These
$f_2$ independent terms are referred to as the DC background terms
\cite{Seidler}. The functions $\Gamma_{1.5}^{(N)}(f_1)$ and
$\Gamma_2^{(N)}(f_1)$ are $\Gamma_{1.5}(f_1)$ and $\Gamma_2(f_1)$
normalized by $\sum_t H(t,f_1)$ and $(\sum_t H(t,f_1))^2$,
respectively. Consideration of $Im\{S_{1.5}(f_2,f_1)\}$ is left for
a long paper.

        We have left Eq. (\ref{theta1}-\ref{secondSC}) in terms of
        general scaling forms since we ascertain that our mean field
        and IRM simulations obey the same scaling forms only with different
        exponents and non-universal constants.

        
        {\em Mean Field Simulation}: We perform $300$ runs of a
        simulation of the mean field RFIM.  We collect data taken from
        $H \in [0,0.00125]$ at $R=0.79788$ ($R_c=0.79788456$) in
        systems with $N = 15 \times 10^6$ spins, and $J=1$. The
        results of our simulation agree with the scaling forms given
        in Eq.  (\ref{theta1}-\ref{secondSC}) with exponents given by
        Table \ref{expo}. See Fig.
        \ref{haarsumplot}-\ref{secondspecplot}.
        
        {\em Infinite Range Model Simulation}: We perform $60$ runs of
        a 3D simulation of the infinite range RFIM.  The data was
        taken from $H \in [1.25,1.88]$ (from the slanted part of the
        hysteresis loop) at $R=2.2$ in system with $N = 400^3$ spins,
        $J_{inf}=\frac{1}{N}$, and $J=1$.  Again, the results agree
        with Eq. (\ref{theta1}-\ref{secondSC}) with exponents given in
        Table \ref{expo}. Refer to Fig.
        \ref{haarsumplot}-\ref{secondspecplot}.
        
        {\em Finite Size Effects}: We check finite size effects in our
        mean field and IRM simulations by looking at higher order
        spectra for various system sizes.  We find that for smaller
        system sizes the high frequency scaling (and flattening due to
        the DC background term) is unchanged for second spectra
        and $1.5$ spectra. However, at low frequency the scaling
        regime of the second spectra and $1.5$ spectra flattens (in MF
        Sim.) or rolls over (in IRM) at frequency $f_1 \simeq
        \frac{1}{T_{max}}$, where $T_{max}$ is the maximum avalanche
        duration.  $T_{max}$ is system size dependent since
        $T_{max}\sim L^z$, where $L=N^{1/d}$ ($d$ = dimension of
        system). For a $N=400^3$ system we find that $T_{max}\simeq
        4300\Delta t$.

\begin{table}
\begin{tabular}{|c||c|c|c|c|}\hline
& $V_{1.5}$ & $V_2$ & $Q_{1.5}$ & $Q_2$ \\ 
        \hline
        MFT & 2  & 2  & 3  &  5 \\
        \hline
        MF Sim. & $1.93\pm0.10$ & $1.92\pm0.12$ & $2.95\pm0.10$ & $4.93\pm0.12$\\
        \hline
        IRM ($d = 3$) & $1.80\pm0.07$ & $1.80\pm0.05$  & $2.73\pm0.06$ & $4.46\pm 0.07$\\
        \hline
\end{tabular}
\caption{\label{expo} We present the values of the exponents: $V_{1.5}$, $V_2$,
$Q_{1.5}$, and $Q_2$ given in Eqs. (\ref{theta1}-\ref{secondSC}) for
mean field theory (MFT), mean field simulation (MF Sim.), and the
infinite range model (IRM) for $d=3$. While the exponents values for
MFT were determined analytically, the exponents for the MF Sim. and IRM
were determined through a non-linear curve fitting of simulation data.}
\end{table}

{\em Discussion}: We first notice that the exponents for our MF
simulation are systematically smaller (by an amount of $1\%$ to $4\%$)
than the MFT exponents, since the MF simulation results were obtained
from a train of finite avalanches while our MFT calculation was for a
single infinite avalanche. The intermittency introduced by a train of
finite avalanches apparently lowers the magnitude of the mean field
exponents since the intermittency effectively adds white
inter-avalanche noise to the intra-avalanche noise seen in the MF
calculation.

For the DC components ($\theta = 0$) of the higher order spectra
(see Fig. \ref{haarsumplot}) we find excellent agreement between MFT
and the MF simulation. In fact the exponents $Q_2$ and $Q_{1.5}$ are
not novel; with the help of \cite{Dahmen} we ascertain the following
exponent relations: $Q_2 = \frac{5-\tau}{\sigma\nu z} - 2$ and
$Q_{1.5} = \frac{1}{\sigma\nu z}+1$.  Using the values given in
\cite{Mehta} for $\tau$ and $1/\sigma\nu z$ we find exact agreement in
MFT: $Q_2 = 5$ and $Q_{1.5} = 3$, and in the IRM in $d=3$ we find $Q_2
= 4.40\pm0.10$ and $Q_{1.5} = 2.72\pm0.03$, in close agreement with
the table above.

        In Fig. (\ref{halfspecplot}-\ref{secondspecplot}) we present
        the $1.5$ spectra and the second spectra.  We determine the
        high frequency power law exponent for the $1.5$ spectra and
        second spectra given by the exponents $V_{1.5}$ and $V_2$
        respectively (see Table \ref{expo}).  Upon subtracting the
        DC background we find that the second spectra, for
        different $f_1$ values, collapses (see Inset of
        Fig. \ref{secondspecplot}). Since the DC background term
        is small for the $1.5$ spectra, we find a collapse without
        subtracting the DC background.
        
        The weak dependence of $1.5$ spectra on $f_1$ and the
        strong fall off of the second spectra at high frequency in
        both the MF Sim. and IRM suggests that the high
        frequency noise comes from the fine structure of large
        avalanches ($T_a>>1/f_1$, $T_a$ is the avalanche duration) and
        not small individual pulses ($T_a\simeq 1/f_1$)\cite{Petta}.
        Through simulation we verify this claim in the MF Sim.
        and IRM. When we subtract all avalanches smaller than
        $T_{max}/4$ from the Barkhausen train and then determine the
        second spectra we find no change in $V_2$. However, when we
        subtract all avalanches larger than $T_{max}/4$ from the
        Barkhausen train, we find that the second spectra flattens and
        that there is an evident separation between the $1.5$ spectra
        curves (i.e. increased $f_1$ dependence). 
        
        Our study of higher order spectra is a powerful tool to
        further our understanding of noise in disordered systems. Our
        mean field results are applicable to a large array of systems,
        in particular systems discussed in \cite{Fisher}. In a future
        long paper we plan to include details of all the
        calculations, and a comparison with experiment results.

\begin{acknowledgments}
        We would like to thank J. Sethna for helpful discussions, and
        we thank M. Kuntz and J. Carpenter for providing the front
        propagation model simulation code.  K.D. and
        A.P.M. acknowledge support from NSF via Grant Nos. DMR
        03-25939(ITR), the Materials Computation Center, through NSF Grant
        No. 03-14279, and IBM which provided the computers that made
        the simulation work possible. M.B.W. acknowledges support from
        NSF via Grant No. DMR 02-40644.  A.P.M. would also like to
        acknowledge the support provided by UIUC through a University
        Fellowship, and K.D. gratefully acknowledges support through
        an A.P. Sloan fellowship.
\end{acknowledgments}

\end{document}